\documentstyle[preprint,aps]{revtex}
\input epsf	
\begin{document}

\tighten

\preprint{gr-qc/9601019 \qquad NSF-ITP-95-165}

\title{Comment on ``Instabilities in Close Neutron Star Binaries"}

\author{
	Douglas M.~Eardley$^1$ and Eric W.~Hirschmann$^2$}
\address{$^1$Institute for Theoretical Physics,
	University of California,
	Santa Barbara, CA 93106-4030\\
	({\tt doug@itp.ucsb.edu})}
\address{$^2$Dept.~of Physics,
	University of California,
	Santa Barbara, CA 93106-9530\\
	({\tt ehirsch@dolphin.physics.ucsb.edu})}

\date{January 10, 1996}

\maketitle


\bigskip

In a recent Physical Review Letter, Wilson and Mathews~\cite{WM}
presented some interesting numerical calculations of a system of
two equally massive neutron stars in strong-field gravity.  In
particular they estimated the innermost stable circular orbit in
their system.  Here we point out a possibly important consequence
of their results:  Their calculated configurations have total
angular momentum $J$ and total mass $M$ too large to form any Kerr
black hole: $J>M^2$.

For a compact binary system in general relativity, the concept
of the innermost stable orbit is not really well defined, because
the orbit shrinks due to gravitational radiation reaction.
However, because the orbit shrinks slowly, an approximate
definition can be given, see \cite{Cook,KWW,BD,CE} for discussion.
Figure 1 displays various calculations for the innermost stable
orbit in the $(J,M)$ plane;  previous results \cite{Cook,KWW,BD,CE}
give
$J\mathrel{\raise.3ex\hbox{$<$\kern-.75em\lower1ex\hbox{$\sim$}}}
M^2$.
In contrast, Wilson and Mathews~\cite{WM} find $J/M\approx1.31$
(point $1'$), well outside the allowed region of final states
($J<M^2$) for Kerr black holes (see {\bf Black Hole Limit}
curve).
\footnote{A slightly tighter limit applies from the Area
Theorem~\cite{Cook}, if the binary system begins as two widely
separated, equally massive Schwarzschild black holes (see
{\bf Area Theorem Limit} curve).  This limit does not apply to
neutron star binaries, and in any case is not much tighter here
than the black hole limit.}

The Wilson-Mathews results therefore imply that the system $1$
must radiate substantial amounts of $J$ after loss of
orbital stability, during merger.  The concomitant loss of
$M$ is governed by the relation $\Delta M = \pi f\Delta J$
where $f$ is the gravity wave frequency~\cite{HF} (assuming
quadrupole emission).  One
expects that $f$ will substantially exceed the final orbital
frequency (about 410 Hz in [1]), since the holes ought to plunge
quickly together after loss of orbital stability.  In Fig.~1,
conjectural evolutions are shown as dotted arrows, labeled by
conjectured frequency $f$? in Hz, normalized to
$M_0=2.90M_{\odot}$.  This additional radiation could be
of considerable importance for the detection and
characterization of these sources by LIGO and
VIRGO~\cite{HF}.

An often-discussed possibility could
be realized here, namely the formation of an excited, nearly
extremal Kerr black hole, radiating copiously at its rigid
rotation frequency $f=\Omega_+/\pi=1/(2\pi M)$.  Note however
that this frequency is ``too high'':
the excess $J$ {\it cannot} be primarily lost in this
way, since such an excited hole would evolve {\it parallel}
to the {\bf Black Hole Limit} in Fig.~1, rather than across it.

Could the Wilson-Mathews approximation scheme be inaccurate
enough to explain the whole transgression into the forbidden
region in Fig.~1?  Pending further supercomputing using full
general relativity, this remains an open possibility, but
seems a bit unlikely in view of recent checks on their
scheme~\cite{CST}.

A longer paper by Wilson, Mathews and Marronetti~\cite{JMP}
provides further details of their methods, and also notes that
$J>M^2$ for their systems.

\global\firstfigfalse
\begin{figure}
\centerline{\epsfysize=5.5in\epsfbox[20 300 592 779]{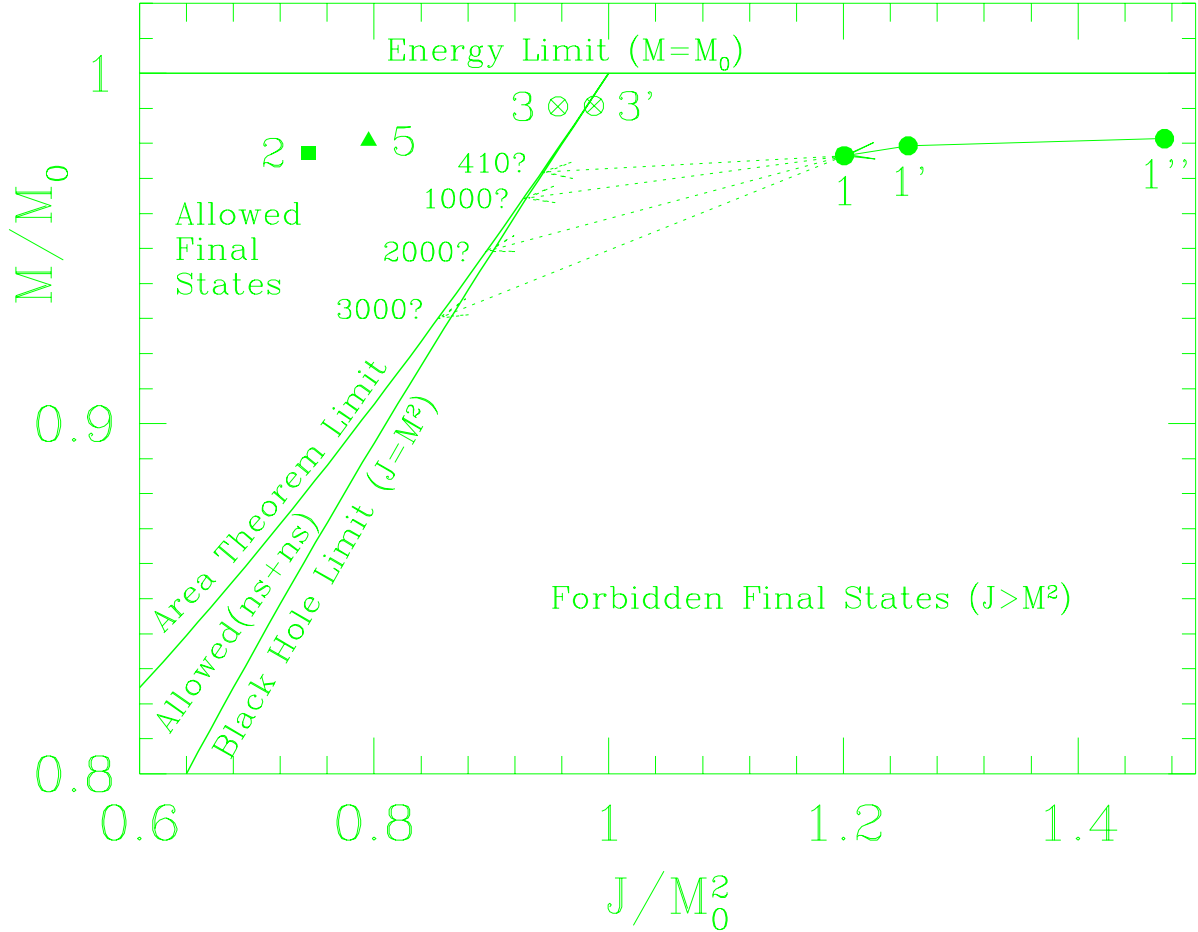}}
\caption{Equal-mass binary systems near the innermost stable circular
orbit, plotted as total system mass $M$ versus angular momentum $J$
in units of initial total system mass $M_0$ at wide separation.
Points labeled $n$ are from reference [$n$]. Points $1$, $1'$, $1"$
are three numerical relativity calculations by Wilson and Mathews [1]
who conclude that $1'$ approximates the innermost stable orbit.  Point
$2$ is an initial value calculation by Cook [2] for same.  Points $3$,
$3'$ are (post)$^2$ calculations for same by Kidder, Will and Wiseman
[3]; $3$ is the static result, while $3'$ includes orbital shrinkage
due to gravitational radiation reaction.  See text.}
\label{fig1}
\end{figure}

\bigskip


\end{document}